\documentclass[final,3p]{elsarticle}
\makeatletter
\def\ps@pprintTitle{%
 \let\@oddhead\@empty
 \let\@evenhead\@empty
 \def\@oddfoot{\centerline{\thepage}}%
 \let\@evenfoot\@oddfoot}
\makeatother
\usepackage{amssymb}
\usepackage{amsmath}
\usepackage{amsthm}
\usepackage{enumerate}
\usepackage{hyperref}
\usepackage{pifont}
\newcommand{\xmark}{\ding{55}}%

\usepackage{lineno}

\journal{Physica A: Statistical Mechanics and its Applications}

\newcommand{\Ex}{\mathop{\bf E\/}}

\newtheorem{lemma}{Lemma}

\newtheorem*{theorem*}{Theorem}
\newtheorem*{lemma*}{Lemma}

\theoremstyle{definition}

\begin{document}

\begin{frontmatter}


%% and the form \Ex\ead[url] for the home page:
%% \Ex\ead{email address}
%% \Ex\ead[url]{home page}

\title{Impact of temporal network structures on the speed of consensus formation in opinion dynamics}


\author{Mingwu Li}
\ead{mingwul2@illinois.edu}
\author{Harry Dankowicz}
\address{Department of Mechanical Science and Engineering, University of Illinois at Urbana-Champaign, Illinois,
USA}

\begin{abstract}
% general background and your motivation
Opinion dynamics on networks has wide applications to empirical and engineered systems and profound prospects in the general study of complex systems. Many efforts have been devoted to understanding how opinion dynamics is affected by network topology. However, human social interactions are best characterized as temporal networks in which ordering of interactions cannot be ignored. Temporal activity patterns including heterogeneous contact strength and interevent times, turnover edge/node dynamics and daily patterns could have significant effects that would not be captured by static aggregate network representations. In this paper, we study the effects of such temporal patterns on the speed of consensus formation in various models of continuous opinion dynamics using three empirical human face-to-face networks from different real-world settings.
We find that static, aggregated networks consistently overestimate the speed of simulated consensus formation while weight heterogeneity associated with frequency of interactions has an inhibitory effect on consensus formation relative to the behavior on unweighted networks. Moreover, the speed of consensus formation is found to be highly sensitive to nodal lifetimes, suggesting that randomization protocols that dramatically alter the distribution of lifetimes cannot be relied upon as reference models. On the other hand, temporal patterns including burstiness of interevent times and the lifetime of edges are found to have insignificant effects on consensus formation.
\end{abstract}


\end{frontmatter}


\section{Introduction}
% introduce opinion dynamics and its background 
Social interactions may have significant influence on decision-making and opinion formation on networks~\cite{social} with implication to political affiliation~ \cite{voter-1,political} and rumor spreading~\cite{rumor}. In social networks, individuals adjust their opinions based on mutual interactions and group discussions. Consensus may result in which all individual hold the same opinion. Alternatively, polarization and fragmentation in which multiple opinion clusters form, could occur as well. The effects of network structure and, particularly, temporal sequencing of interactions on such opinion dynamics are the concern of this paper.

% talk discrete models and continuous opinion models
Opinion dynamics on static network topologies have been explored in terms of iterated maps applied to discrete~\cite{voter-1,voter-2,tutorial} or continuous representations of opinions~\cite{cont-suvey}. The maps defined by Deffuant et al~\cite{Deff} and Hegselmann and Krause~\cite{HK} account for topology by, in the first case, randomly selecting an edge with probability proportional to its network weight and updating the corresponding pair of nodes or, in the second case, updating all nodes simultaneously using weighted averages of neighboring opinions~\cite{tutorial}. The Deffuant model has attracted a large amount of attention of researchers from communities of sociophysics and network science. It was first applied to complete graphs such as square lattices~\cite{Deff} and subsequently to generative networks by Barab\'{a}si-Albert, Erd\H{o}s-R\'{e}nyi, and Watts-Strogatz mechanisms~\cite{B-A,adaptive,small-world,W-S}. It was also modified to adaptive networks where the breaking and rewiring processes of links are introduced~\cite{adaptive,political}. Recently, the Deffuant model was applied to empirical data sets~\cite{mason}. Various generalizations include smoothing of so-called confidence bounds to avoid discontinuous jumps in opinion differences~\cite{deffuant2002can}, as well as heterogeneous and time-dependent confidence bounds~\cite{time-dependent,heterogeneous}.

% here we focus on temporal networks
Unti recently, few studies have considered the opinion dynamics on temporal networks in which the ordering of interaction is properly accounted for~\cite{temp-2012,temp-2015}. Temporal networks are a natural framework for describing human face-to-face interactions because the spatial proximity between two individuals is time-varying. It has been demonstrated that temporal patterns like contact orders~\cite{causa}, burstiness~\cite{null-2}, and lifetimes~\cite{ongoing,lifetime} could have significant effects on dynamics on networks, and may yield totally different results from static network representations. We are aware of only limited work on voter models (with binary opinions) on temporal networks~\cite{temp-voter-1,temp-voter-2,temp-2015}. Even less attention has been put on dynamic models of continuous opinions on temporal networks, with the sole exception of a recent paper that  explored continuous opinion dynamics on synthetic activity-driven networks~\cite{li2018opinion}. Motivated by this, we use three publicly-available empirical data sets representing various examples of human proximity interactions to investigate how temporal patterns affect consensus formation speeds in continuous opinion dynamics.
%[UNDO: SAY SOMETHING ABOUT HUMAN INTERACTION NETWORKS, HIGHLIGHTING ITS NATURE OF TEMPORAL NETWORKS. GIVE MORE DETAILED DISCUSSION REGARDING THE EFFECTS OF SPEED-UP AND SLOW-DOWN, ALSO MENTIONING OUR RECENT WORK]

The rest of this paper is organized as follows. In section~\ref{sec:models}, we review the Deffuant and Hegselmann/Krause models and generalize these to temporal networks. We choose to characterize the speed of consensus formation in terms of the dynamics of the expectation of the mean square opinion difference across the network. In section~\ref{sec:empirical networks}, we give basic information of three empirical networks and highlight their temporal structures. In section~\ref{sec:comp-static}, we compare the consensus speed on aggregated unweighted and weighted networks to probe the effects of edge weights. Comparison of the speed of consensus formation on temporal networks and the corresponding aggregated weighted networks is presented to highlight the potential issues of static models. In section~\ref{sec:impact-temp}, we carefully examine the effects of temporal patterns on the speed of consensus formation using several different reference models. A brief summary and discussion is considered in the concluding section~\ref{sec:conclusion}.

%===================================================%

%==================================================%
\section{Models of opinion dynamics}
\label{sec:models}
%---------------------------------%

We begin by defining a collection of iterated maps whose finite-duration dynamics model changes in the distribution of opinions across a network described in terms of a static or time-dependent topology.

\subsection{Static networks}
\label{sec:static networks}
Consider first a static, weighted, connected graph on a network with $N$ nodes, $E$ edges, and Laplacian $L$. Let $L^{(ij)}$ denote the Laplacian associated with the unweighted graph obtained by removing all edges but that connecting nodes $i$ and $j$, such that $L^{(ij)}=0$ if no such edge exists in the original graph. It follows that $L^{(ij)}$ is obtained by setting the weight of the edge connecting nodes $i$ and $j$ to $1$ and the weights of all other edges to $0$.

Let $x=[x_1,\ldots,x_N]^\top\in\mathbb{R}^N$ denote the vector of \emph{opinions} across the network. We assume that updates to $x$ are governed by a linear, possibly non-autonomous, discrete map of the form
\begin{equation}
    x(k+1) = T_k\,x(k).
\end{equation}
If $n$ denotes the discrete duration of the corresponding dynamics, then
\begin{equation}
\label{eq:transition}
    x(n) = T_{n0}\,x(0),
\end{equation}
where $T_{n0}=T_{n-1}...T_0$ is the corresponding \emph{opinion transition matrix} and $x(0)$ is the \emph{initial opinion} across the network.

Different update schemes yield different forms of the opinion transition matrix. In the original Deffuant model~\cite{Deff}, opinions are updated sequentially in pairs by selecting a network edge $(ij)$ randomly at each iteration. Here, if the distance $|x_i(k)-x_j(k)|$ between the corresponding two nodal opinions exceeds a predefined \emph{confidence bound} $\epsilon$, then $x(k+1)=x(k)$. In contrast, when $|x_i(k)-x_j(k)|\leq\epsilon$,
\begin{equation}
\label{eq:update}
    x{(k+1)} = (I-\mu L^{(ij)})x(k),
\end{equation}
where the \emph{convergence parameter} $\mu$ represents the tendency of nodes to align their opinions with those of their neighbors. In this paper, we assume that $\mu \in (0,1)$. In this case,
\begin{equation}
\label{eq:dist-up}
|x_i{(k+1)}-x_j{(k+1)}|=|1-2\mu|\cdot|x_i{(k)}-x_j{(k)}|
\end{equation}
shows that the opinion difference of the two nodes is reduced as a result of the interaction. 
%(commonly set to $0.5$~\cite{adaptive, political})

If we ignore the confidence bound $\epsilon$, then it follows from Eq.~\eqref{eq:update} that
\begin{equation}
    T_{n0} = \prod_{k=0}^{n-1} \left(I-\mu L^{(i_kj_k)}\right).
\end{equation}
Since the selected edge sequence is a random sequence with replacement, the corresponding sequence $\{L^{(i_kj_k)}\}_{k=0}^{n-1}$ of Laplacian matrices is also a random sequence with replacement. If we assume that the initial opinion is selected from some probability distribution that is independent of the distribution of Laplacian matrices, then the expected value
\begin{equation}
\label{eq:stat-mean}
    \Ex \left[x{(n)}\right] =  \prod_{k=0}^{n-1} \left(I-\mu \Ex \left[L^{(i_kj_k)}\right] \right) \Ex \left[x{(0)}\right] = \left(I-\mu \overline{L} \right)^n \Ex \left[x{(0)}\right],
\end{equation}
where $\Ex\left[L^{(i_kj_k)}\right]=\overline{L}$, for $k=0,\ldots,n-1$, due to replacement. The expectation of the opinion transition matrix then equals
\begin{equation}
    \overline{T}_{n0}=\left(I-\mu \overline{L} \right)^n.
\end{equation}
If the probability of an edge being selected is proportional to its weight, then
\begin{equation}
    \overline{L}=L/\sum w_{(ij)},
\end{equation}
where $w_{(ij)}$ denotes the weight of the edge connecting nodes $i$ and $j$. In the special case that the network is unweighted (i.e., if $w_{(ij)}$ equals either $1$ or $0$), then $\overline{L}=L/E$. In either case, it follows from the appendix that $\overline{T}_{n0}$ is a row-stochastic matrix.

In the Hegselmann-Krause model~\cite{HK,sirbu2017opinion}, the opinions of all nodes in a static network are updated simultaneously according to the deterministic dynamics
\begin{equation}
x_i(k+1) = \frac{\sum_{j:|x_i(k)-x_j(k)|\leq\epsilon}a_{ij}x_j(k)}{\sum_{j:|x_i(k)-x_j(k)|\leq \epsilon}a_{ij}} \quad \text{if}\quad {\sum_{j:|x_i(k)-x_j(k)|\leq\epsilon}a_{ij}}>0,
\end{equation}
and $x_i(k+1)=x_i(k)$ otherwise. Here, the $a_{ij}$'s are the entries of the network adjacency matrix. While the Deffuant model captures pairwise interactions within large populations, the Hegselmann-Krause model describes situations where opinion dynamics are a result of group conversations, for example, in formal meetings~\cite{sirbu2017opinion}.

In the remainder of this section, we propose four generalizations of these models to finite-duration dynamics on \emph{time-dependent} networks.  In later sections, we proceed to analyze the effects of temporal patterns on the opinion dynamics obtained by applying these generalizations to publicly-available network data sets. Since we focus on the speed of consensus formation, we simplify our analysis by ignoring the confidence bound $\epsilon$.

\subsection{Time-dependent networks}
\label{sec:timedep networks}

Consider a sequence $G=\{G_l\}$ of $m$ unweighted graphs on a network with $N$ nodes, and let $\{L_l\}$ denote the corresponding sequence of  Laplacians. Let $e_l$ denote the number of edges in $G_l$ and let $M=\sum_l e_l$. Let $G^u$ denote a time-aggregated, unweighted, static topology on the same network with Laplacian $L^u$, such that every edge in $G^u$ occurs at least once in the sequence $G$, and every edge in $G_l$, for $l=1,\ldots,m$, occurs in $G^u$. Although individual graphs in the sequence $G$ may be disconnected, we assume that $G^u$ is connected. Let $G^w$ denote a time-aggregated, weighted, static topology on the same network with Laplacian $L^w$, whose edges are identical to those in $G^u$, but with nontrivial weights corresponding to the number of occurrences of each edge in the sequence $G$. It follows that $G^w$ is also connected.

As a baseline approach, we first consider the application of the stochastic Deffuant model \eqref{eq:update} to the static networks $G^u$ and $G^w$ with likelihood of selection of an individual edge proportional to its weight. The results from the previous section then apply with $L$ equal to $L^u$ and $L^w$, respectively. In either case, the expected transition matrices are row-stochastic matrices. For comparison with the stochastic update model described next, we let $n=M$.

Alternatively, we retain the ordering inherent in the original sequence $G$ by first applying stochastic Deffuant updates to the graph $G_1$ as many times as the number of edges in $G_1$, then applying stochastic Deffuant updates to the graph $G_2$ as many times as the number of edges in $G_2$, and so on, until the sequence is exhausted. It follows that
\begin{equation}
\label{eq:asynchro}
    T_{M0}=\prod_{l=1}^m\prod_{k=0}^{e_l-1}\left(I-\mu L_l^{(i_{l,k}j_{l,k})}\right)\Rightarrow    \overline{T}_{M0}=\prod_{l=1}^m\left(I-\mu \frac{L_l}{e_l}\right)^{e_l},
\end{equation}
From the appendix we conclude that the matrix $\overline{T}_{M0}$ is again a row-stochastic matrix.

Finally (cf.~\cite{li2018opinion}), we consider the application of deterministic synchronous updates to all nodes in $G_1$, followed by the application of deterministic synchronous updates to all nodes in $G_2$, and so on, until the sequence is exhausted. In this case, we let $\mathcal{N}_{l,i}$ denote the set of neighbors of the $i$th node in the $l$th graph and define
\begin{equation}
\label{eq:synchro}
    x_i{(l+1)} = x_i{(l)}+ \frac{\mu}{|\mathcal{N}_{l,i}|} \sum_{j\in \mathcal{N}_{l,i}} \left(x_j{(l)}-x_i{(l)}\right)
\end{equation}
if $|\mathcal{N}_{l,i}|\neq 0$, and $x_i{(l+1)}=x_i{(l)}$ otherwise, for $i=1,\ldots,N$. In this case,
\begin{equation}
    T_{m0}=\prod_{l=1}^m\left(I-\mu w_lL_l\right),
\end{equation}
where $w_l$ is a diagonal matrix with $i$th diagonal element $1/|\mathcal{N}_{l,i}|$ if $|\mathcal{N}_{l,i}|\neq 0$ and $0$ otherwise. It again follows from the analysis in the appendix that $T_{m0}$ is a row-stochastic matrix.

We note that when $\mu=1$, each synchronous update \eqref{eq:synchro} coincides with the Hegselmann-Krause model without confidence bound. In~\cite{li2018opinion}, the generalization in \eqref{eq:synchro} is applied to activity-driven networks, in which each consecutive graph is generated synthetically from a reference collection of unconnected nodes by ``activating'' individual nodes with some probability according to a given distribution and connecting each active node to a finite number of other randomly chosen nodes. In contrast, in our study, we focus on applying the update model in \eqref{eq:synchro} to empirical network data in order to characterize the corresponding temporal structure. Specifically, in Section~\ref{sec:impact-temp}, we apply each of the generalized update models to the sequence of graphs $\{G_l\}_{l=1}^m$ obtained from individual time stamps associated with start times of interactions in three empirical data sets and their randomized reference networks. 
%We thus obtain
%\begin{equation}
%    T_s^u=\left(I-\mu\frac{L^u}{E}\right)^M,\,T_s^w=\left(I-\mu\frac{L^w}{\sum w_{(ij)}}\right)^M,\,T_t^{as}=\prod_{l=1}^m\left(I-\mu \frac{L_l}{e_l}\right)^{e_l},\,\text{and}\,T_t^s=\prod_{l=1}^m\left(I-\mu w_lL_l\right).
%\end{equation}

\subsection{Consensus formation}
\label{sec:consensus}

In the special case that all components of the initial opinion vector $x(0)$ are governed by identical independent probability distributions with mean $\overline{m}$, it follows from row-stochasticity that
\begin{equation}
\label{eq:expectation}
    \Ex\left[x_i{(n)}\right]= \sum_{j=1}^N \overline{T}_{ij} \Ex\left[x_j{(0)}\right] = \overline{m}\sum_{j=1}^N \overline{T}_{ij} = \overline{m},\, \forall i\in\{1,...,N\},
\end{equation}
where $\overline{T}_{ij}$ denotes the $(i,j)$th component of any of the opinion transition matrices or, in the case of stochastic updates, their expectations. The expected opinion thus remains equal to the expectation of the initial opinion consistent with the possibility of consensus formation. 

The dynamics of consensus formation across the network may be quantified in terms of the \textit{mean square opinion distance} between two different nodes~\cite{accel}
\begin{equation}
    d(k) = \frac{2}{N(N-1)}\sum_{i=1}^N\sum_{j=1}^{i-1}\left(x_i{(k)}-x_j{(k)}\right)^2
\end{equation}
with corresponding expectation
\begin{align}
\label{eq:dist}
    \Ex(d(k)) =  \frac{2}{N(N-1)}\sum_{i=1}^N\sum_{j=1}^{i-1}\Ex\left[\left(x_i{(k)}-x_j{(k)}\right)^2\right].
\end{align}
In the special case of the deterministic update model in \eqref{eq:synchro},
\begin{align}
\label{eq:dist-ij}
    &\Ex\left[\left(x_i{(k)}-x_j{(k)}\right)^2\right] = \Ex\left[\left(\sum_{p=1}^N (T_{k0,ip}-T_{k0,jp}) x_p{(0)}\right)^2\right]\nonumber\\
    & = \sum_{p=1}^N \left(T_{k0,ip}-T_{k0,jp}\right)^2 \widehat{m}+\sum_{q=1}^N \sum_{p=1}^{q-1} 2(T_{k0,iq}-T_{k0,jq})(T_{k0,ip}-T_{k0,jp}) \overline{m}^2,
\end{align}
where $\widehat{m}$ represents the second moment of the distribution of initial opinions. No such explicit expressions are available to evaluate $\Ex(d(k))$ for the stochastic update models. This must instead be obtained by appropriate numerical sampling and simulation.

\section{Empirical networks}
\label{sec:empirical networks}

%-------------------------------%
\subsection{Basic information and statistics}
\label{sec:emp-basic}
We apply our analysis to three publicly-available, empirical, time-dependent network data sets, here labeled ``Conference''~\cite{Workplace}, ``High school''~\cite{Highschool}, and ``Hospital''~\cite{Hospital}. These datasets have different spatial and temporal characteristics. For example, the ``Conference'' data set corresponds to a homogeneous topology, while the ``High-school'' data set exhibits community structure. Temporal patterns in the ``High-school'' and ``Conference'' data sets follow relatively strict and repetitive schedules, while activities in the ``Hospital'' network data set are less regular.

All three data sets were collected from face-to-face proximity contacts of individuals with wearable sensors. Data is recorded as contact sequences, with each entry identifying two interacting nodes and their contact timestamp. The sampling resolution in each data set is 20 seconds. From a sequence of contacts, we infer a sequence of interactions. Specifically, any pair of contacts on an edge with inter-contact times equal to the sampling resolution are condensed into a single interaction. In our analysis, each data set is associated with a sequence of graphs obtained from the start times of such interactions.

In the ``Conference'' network, we observed that one node has only one neighbor, while all remaining nodes have more than 10 neighbors. Coordination dynamics and diffusive dynamics are highly sensitive to the presence of outliers with significantly smaller degree than the rest of the nodes~\cite{longlast,accel}. For this reason, the outlier node and related links in the ``Conference'' network were deleted in refs.~\cite{longlast,accel}. Due to the close relations between consensus dynamics and coordination~\cite{multi-agent-05,syn}, this outlier node and the corresponding network edge were therefore excluded from our analysis as well.

Table~\ref{basic_stat} lists basic information about the three data sets before and after elimination of the outlier node in the ``Conference'' data set and condensation of contacts into interactions.

\begin{table}[!ht]
\begin{center}
\begin{tabular}{ |c|c c c c c| } 
 \hline
   & $N$ & $E$ & $M$ & $\Delta T$ & $m$\\ 
 \hline
 ``Conference'' & 112 (113) & 2195 (2196) & 9864 (20818) & 2 days & 3659 (5246)\\
 ``High school'' & 180 & 2220 & 19774 (45047) & 7 days & 8468 (11273)\\ 
 ``Hospital'' & 75 & 1139 & 14037 (32424) & 4 days & 6821 (9453)\\ 
 \hline
\end{tabular}
\caption{Basic statistics for the three empirical data sets. Values enclosed within parentheses represent original data before elimination of the outlier network node in the ``Conference'' data set and condensation of contacts into interactions as described in section~\ref{sec:emp-basic}. Here, $N$ and $E$ are the numbers of network nodes and edges, respectively, aggregated over the entire sampling time $\Delta T$. Furthermore, $M$ is the number of interactions/contacts and $m$ is the number of corresponding timestamps in each empirical data set.}
\label{basic_stat}
\end{center}
\end{table}

%------------------------------%
\subsection{Temporal patterns on nodes and edges}
\label{sec:emp-patterns}
\begin{figure}[!ht]
\centering
\includegraphics[width=0.95\textwidth]{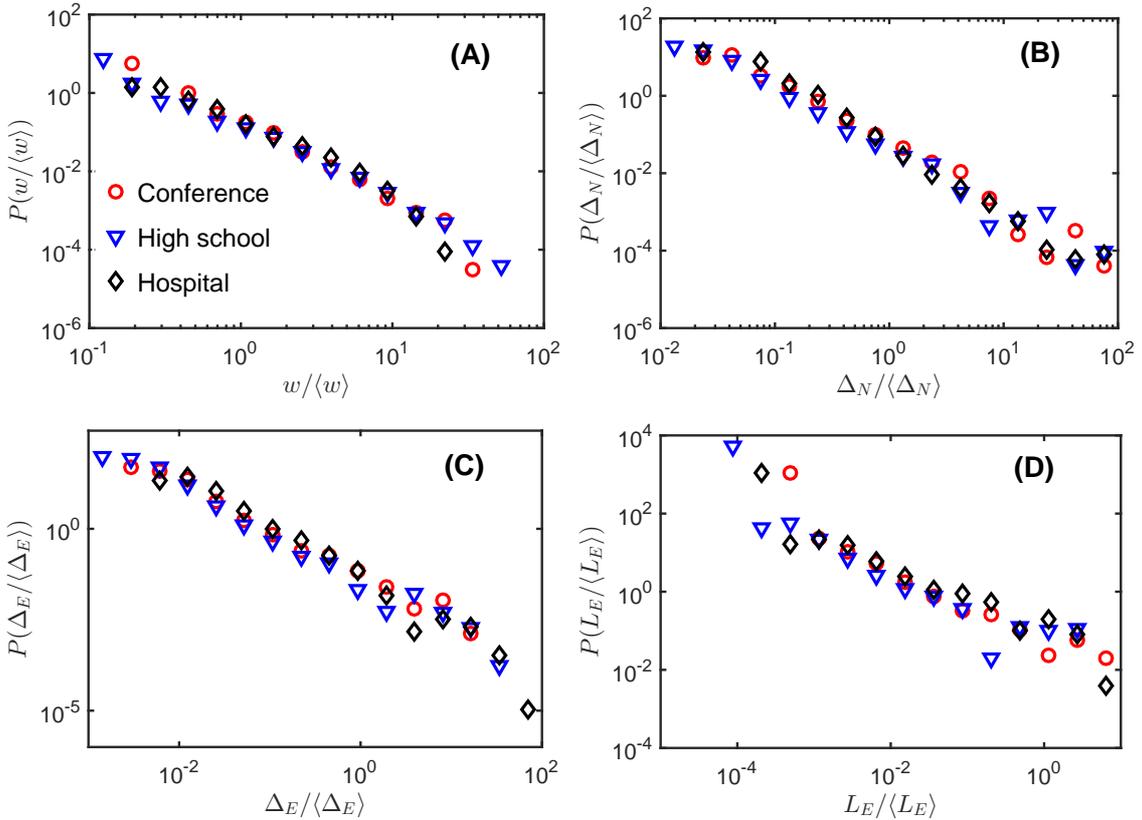}%
\caption{Distributions of (A) edge weights, $w$, (B) node IETs, $\Delta_N$, (C) edge IETs, $\Delta_E$, and (D) edge lifespans, $L_E$, for the three empirical networks. Bin widths are uniform in log space in all cases.}
\label{temp_struct}
\end{figure}

In time-aggregated representations of temporal networks, the weight
of an edge is given by the total number of interactions between the pair of connected nodes. The distribution of edge weights for the three empirical networks is presented in Fig.~\ref{temp_struct}(A). We observe scale-free distributions implying heterogeneous strengths of ties between nodes in each network.

We explore the temporal characteristics of each network by defining node/edge interevent times (IETs) as the time gaps between start times of consecutive interactions associated with individual nodes or edges. Figures~\ref{temp_struct}(B) and \ref{temp_struct}(C) show the corresponding distributions across all nodes for each of the three data sets. As with the edge weights, we find that both node and edge IETs have broad, fat-tailed distributions, indicating bursty temporal structures~\cite{temp-2015}.

A further temporal characterization is afforded by the distributions of edge lifespans for the three empirical networks shown in Fig.~\ref{temp_struct}(D). Here, the lifespan of an edge is the duration between the start times of the first and last interactions occurring on that edge. As seen in the figure, all three networks display heterogeneous distributions of edge lifespans. Such heterogeneity has also been observed in human telecommunication networks~\cite{mobile} and insect proximity networks~\cite{ant}, and suggests that the so-called \emph{turnover picture}  gives a better description of the edge dynamics than the \emph{ongoing picture}~\cite{ongoing}.

\begin{figure}[!ht]
\centering
\includegraphics[width=0.70\textwidth]{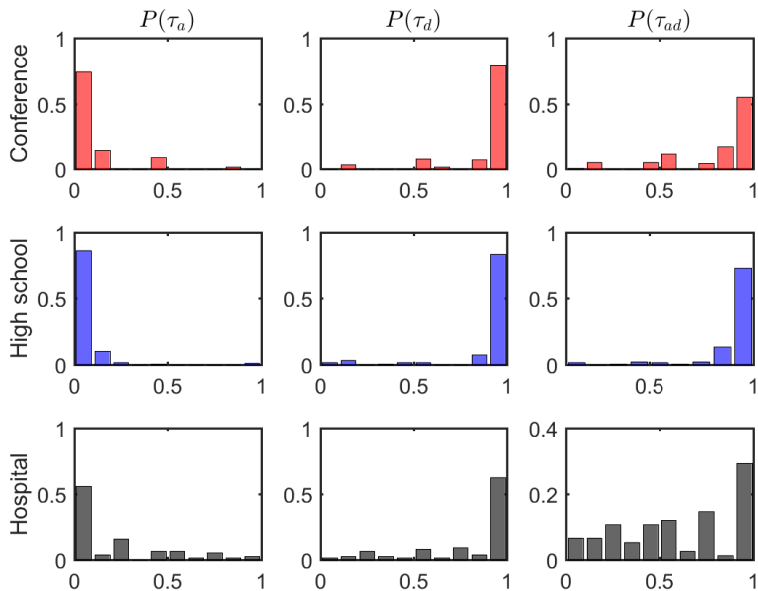}
\caption{Distributions of normalized activation times, $\tau_a$, deactivation times, $\tau_d$, and lifespans, $\tau_{ad}$, of nodes for the three empirical networks. In all cases, bin sizes are set to 0.1.}
\label{node_lifetime}
\end{figure}

A more nuanced description applies to the node lifespans, defined by the duration between the start times of the first (activation) and last (deactivation) interactions associated with each node. In the turnover picture, we expect nodes to enter and leave the networks dynamically. In contrast, in the ongoing picture, most nodes are present and active throughout the sampling time. Following~\cite{lifetime}, we define the normalized activation time, deactivation time, and lifespan of a node as follows
\begin{equation}
    \tau_a=\frac{t_a}{\Delta T};\quad \tau_d=\frac{t_d}{\Delta T};\quad \tau_{ad}=\frac{t_d-t_a}{\Delta T},
\end{equation}
where $t_a$ and $t_d$ are the activation and deactivation times (relative to the start of sampling) and $\Delta T$ is the sampling time. As can be seen in Fig.~\ref{node_lifetime}, the ``High school'' network is best characterized by the ongoing picture. In contrast, ``Hospital'' exhibits significant turnover dynamics with a majority of nodes entering and leaving the network at times interior to the sampling window. The ``Conference'' network falls somewhere between these two extremes, with about half of all nodes not continuously present.

%=====================================================%
\section{Numerical results}
\label{sec:numerical results}

We proceed to analyze the dynamics of consensus formation using the expectation $\Ex(d(k))$ on each of the three empirical datasets using the four distinct update models introduced in Section~\ref{sec:models}. Throughout, we assume that the components of the initial opinion vector are governed by a uniform distribution on $[0,1]$ with mean $\overline{m}=1/2$ and second moment $\widehat{m}=1/3$.

\subsection{Initial characterization}
\label{sec:comp-static}

\begin{figure}[!ht]
\centering
\includegraphics[width=1.0\textwidth]{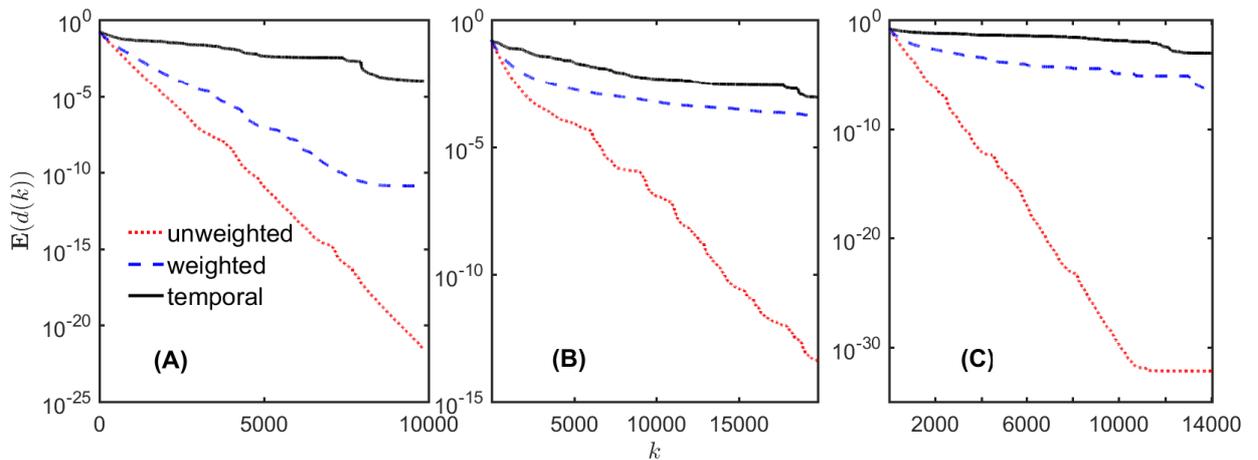}
\caption{Evolution of $\Ex(d(k))$ in the case of stochastic Deffuant updates on $G^u$ (dot line), $G^w$ (dashed line) and sequentially on $\{G_l\}$ (solid line) with $\mu=0.8$. (A) ``Conference'', (B) ``High school'' and (C) ``Hospital''.}
\label{fig:static-temp-traj}
\end{figure}

Following the methodology in Section~\ref{sec:timedep networks}, we first apply the stochastic Deffuant model \eqref{eq:update} to the aggregated topologies $G^u$ and $G^w$ to explore the influence of edge weights. The corresponding results in Fig.~\ref{fig:static-temp-traj} were obtained by averaging $d(k)$ at each of $M$ steps over $100$ different simulations with $\mu=0.8$. The figure shows that $\Ex(d(k))$ decreases with $k$ in both static representations, indicative of a trend toward consensus. The average mean square opinion distance on the unweighted network is much smaller than on the weighted network, suggesting that weight heterogeneity inhibits consensus formation. This inhibitory effect can be explained by the weakness of strong ties~\cite{strong_tie_1, strong_tie_2}, as nodes connected by strong ties are less likely to also make contact with other nodes. More careful observation suggests the $\Ex(d(k))$ on the unweighted network decays exponentially with $k$. In comparison, as shown by the solid curves in Fig.~\ref{fig:static-temp-traj}, consensus formation appears to be significantly slower for opinion dynamics modeled by the sequential application of stochastic Deffuant updates to each of the graphs $G_l$ as described in Section~\ref{sec:timedep networks}. We attribute this observation to the preservation of temporal structures such as burstiness of interevent times and lifetimes of edges and nodes, all of which are destroyed in the aggregated networks $G^u$ and $G^w$. We will carefully examine how temporal structures affect the speed of consensus formation in the next subsection.

\begin{figure}[!ht]
\centering
\includegraphics[width=1.0\textwidth]{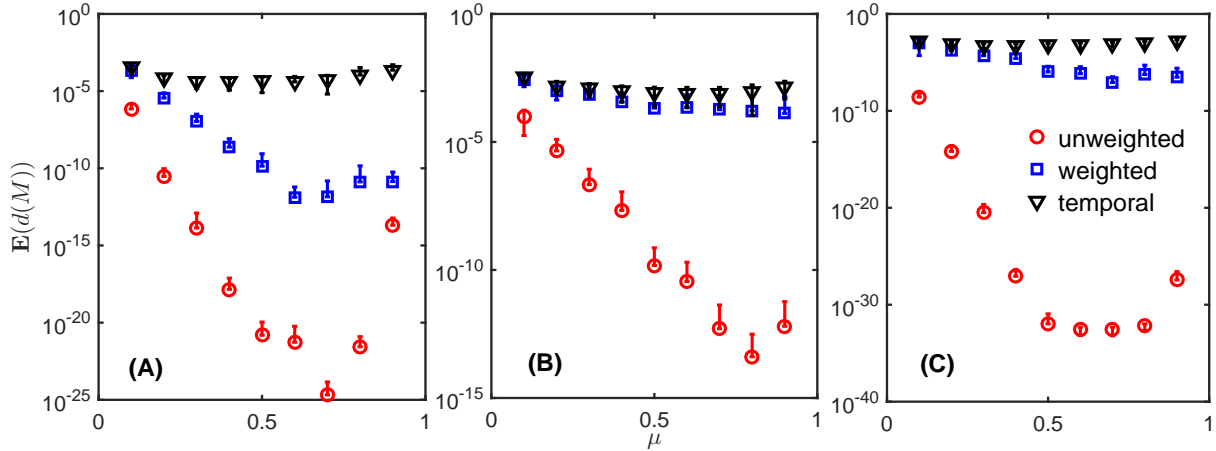}
\caption{Variations in $\Ex(d(M))$ with respect to $\mu$ for opinion dynamics governed by stochastic Deffuant updates applied to $G^u$, $G^w$, and $\{G_l\}$. (A) ``Conference'', (B) ``High school'' and (C) ``Hospital''. Error bars represent averages $\pm$ one standard deviation for 100 realizations. In cases where the average minus one standard deviation results in a negative number, we omit the lower half of the corresponding error bar.}
\label{fig:static-temp-end}
\end{figure}

Figure~\ref{fig:static-temp-end} shows the dependence of $\Ex(d(M))$ on the convergence parameter $\mu$. While the effect of variations in $\mu$ is insignificant in the case of the application of stochastic Deffuant updates to the sequence $\{G_l\}$, dramatic variations are seen in the averaged terminal mean square distance computed on the two time-aggregated topologies. We again see a possible influence from edge weight heterogeneity in the different orders of magnitude of $\Ex(d(M))$ between the results obtained on $G^u$ and $G^w$, respectively, and a clear distinction between results obtained by time-aggregation versus preservation of temporal structure. These differences are particularly strongly accentuated for $\mu$ around $0.7$.

%In the limit case $\mu=1$, we do not expect consensus formation because opinions are simply exchanged (see Eq.~\eqref{eq:update}), which explains the above increment of opinion distance when $\mu\to1$.

%====================================================%
\subsection{Impact of temporal patterns}
\label{sec:impact-temp}

In order to study the effects of temporal patterns on the speed of consensus formation, we make comparisons with network realizations of randomized reference models that destroy the temporal patterns but preserve aggregated topology. As summarized in Table~\ref{struct_null}, each successive reference model listed below destroys additional temporal structure.
\begin{itemize}
    \item Permuted-time model with preserved edge lifetime (PTE)~\cite{lifetime}: Start times of interactions are randomly permuted if the activation and deactivation times of each edge are preserved after permutation. Permutations that change edge lifetimes are rejected. Burstiness of edge IETs is destroyed by PTE.
    \item Permuted-time model with preserved node lifetime (PTN)~\cite{Tim,lifetime}: Start times of interactions are randomly permuted if the activation and deactivation times of each node are preserved after permutation. Permutations that change node lifetimes are rejected. Note that node lifetimes are automatically preserved in PTE while PTN changes edge lifetimes. Burstiness of edge IETs and of node IETs is destroyed by PTN. 
    \item Permuted-time model (PT)~\cite{null-1,null-2}: Start times of interactions are randomly permuted without constraint. Neither node nor edge lifetimes are preserved. Burstiness of edge IETs and of node IETs are destroyed by PT.
    \item Randomized-time model (RT)~\cite{null-1,null-2}: Start times of interactions are replaced by random numbers generated from a uniform distribution over the same sampling window as the corresponding empirical network. Neither node nor edge lifetimes are preserved. Burstiness of edge IETs and of node IETs are destroyed by RT. RT further destroys daily patterns.
\end{itemize}

\begin{table}[!ht]
\begin{center}
\begin{tabular}{|c |c c c c c c c c c c c|}
\hline
 & $t_E^a$ & $t_E^d$ & $L_E$ & $\langle{\Delta}_E\rangle$ & $t_N^a$ & $t_N^d$ & $L_N$ & $\langle{\Delta}_N\rangle$ & D & $m$ & B\\
 \hline
 PTE & $\checkmark$ & $\checkmark$ & $\checkmark$ & $\checkmark$ & $\checkmark$ & $\checkmark$ & $\checkmark$ & $\checkmark$ & $\checkmark$ & $\checkmark$ & \xmark\\
 PTN & \xmark & \xmark & \xmark & \xmark & $\checkmark$ & $\checkmark$ & $\checkmark$ & $\checkmark$ & $\checkmark$& $\checkmark$ & \xmark\\
 PT & \xmark & \xmark & \xmark & \xmark & \xmark & \xmark & \xmark & \xmark &  $\checkmark$ & $\checkmark$ & \xmark\\
 RT & \xmark & \xmark & \xmark & \xmark & \xmark & \xmark & \xmark & \xmark & \xmark & \xmark & \xmark\\
\hline
\end{tabular}
\caption{Temporal structures preserved ($\checkmark$) or destroyed (\xmark) by different reference models. $t_E^a$: edge activation times; $t_E^d$: edge deactivation times; $L_E$: edge lifetimes; $\langle{\Delta}_E\rangle$: mean IET of edges; $t_N^a$: node activation times; $t_N^d$: node deactivation times; $L_N$: node lifetimes;  $\langle{\Delta}_N\rangle$: mean IET of nodes; D: daily patterns; $m$: number of timestamps (or number of graphs in graph sequence); B: bursty IETs.}
\label{struct_null}
\end{center}
\end{table}

We can compare simulation results on empirical networks and their PTE networks to study the effects of bursty edge IETs. Further, comparison of results on PTE and PTN can be used to probe the effects of bursty node IETs and edge lifetimes. In addition, comparison of results on PTN and PT highlight the effects of node lifetimes. Finally, we can compare results on PT and RT to reveal the effects of daily patterns and weekly patterns. We note that although PT has been widely used to study the effects of burstiness in temporal networks, it assumes that edges and nodes are described by the ongoing picture, which does not always hold in empirical networks. The potential bias of PT has been systematically explored in ref.~\cite{lifetime}. In this study, we generate 10 realizations per empirical network and reference model.

\begin{figure}[!ht]
\centering
\includegraphics[width=1.0\textwidth]{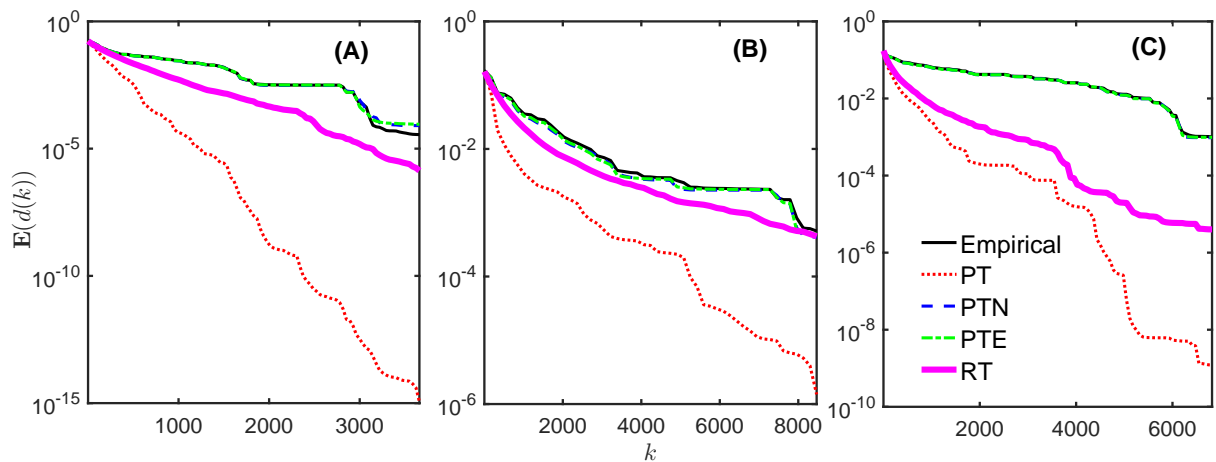}
\caption{Evolution of $\Ex(d(k))$ with $k$ on the empirical networks and their corresponding reference models obtained by deterministic update model \eqref{eq:synchro} with $\mu=0.8$. (A) ``Conference'', (B) ``High school'' and (C) ``Hospital''. The curves for the reference models are averages over 10 realizations.}
\label{fig:temp-ref-sy-traj}
\end{figure}

For the deterministic update model \eqref{eq:synchro}, $\Ex(d(k))$ is given by \eqref{eq:dist} and \eqref{eq:dist-ij}.
Figure~\ref{fig:temp-ref-sy-traj} shows variations in $\Ex(d(k))$ with increasing $k$ in the case that $\mu=0.8$. We observe that the three empirical networks and their PTE and PTN reference models yield close results, suggesting that burstiness of IETs and edge lifetimes have no significant effects on the speed of consensus formation. In contrast, changes to node lifetimes in PT result in significantly smaller expected values of the mean square distance, likely associated with an extension of node lifetimes to the sampling window by this reference model. RT yields slower decay compared with PT. When RT is applied, $m$ typically increases significantly and approaches $M$. Since the total number of interactions is preserved after randomization, the number of edges in each graph will be reduced, typically to $1$. This reduction of interaction strength results in the slower decay of RT. Indeed, even if the timestamps of interactions are randomly selected only from daytime, $\Ex(d(k))$ decays more slowly for RT than for PT. Thus, we cannot probe the effects of daily patterns by comparing simulation results using the synchronous update model on PT and RT.

\begin{figure}[!ht]
\centering
\includegraphics[width=1.0\textwidth]{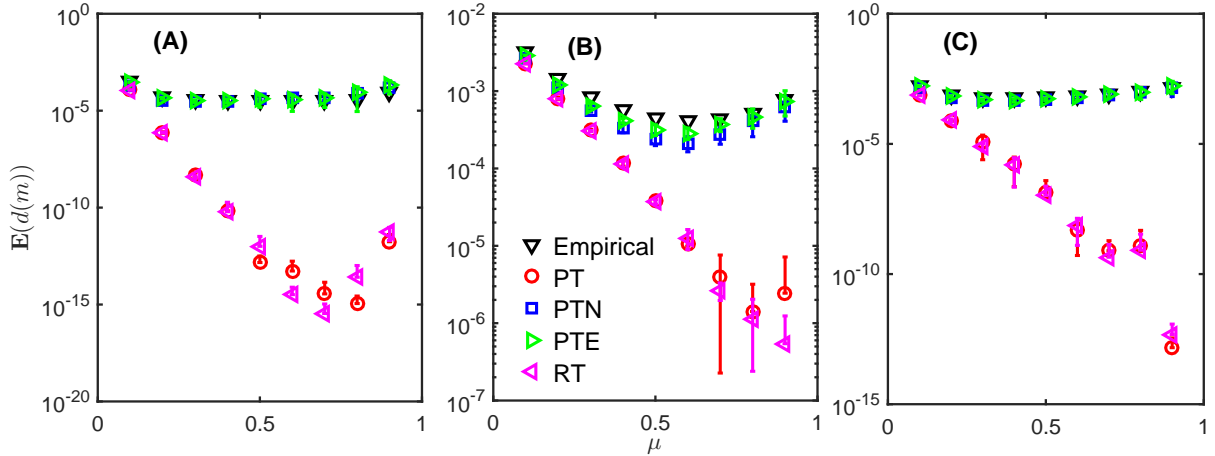}
\caption{The dependence of $\Ex(d(m))$ on $\mu$ for each empirical network and the corresponding reference models using the synchronous update model \eqref{eq:synchro}. (A) ``Conference'', (B) ``High school'', and (C) ``Hospital''. Results for the PTE, PTN, PT and RT reference models are presented using error bar that represent averages $\pm$ one standard deviation over 10 realizations. In the case that the average minus one standard deviation results in a negative number, we omit the lower half of the error bar. It is noted that $m$ is changed by RT. The new value of $m$ has been used here to compute expected terminal mean square distance.}
\label{fig:temp-ref-sy-end}
\end{figure}

The dependence of $\Ex(d(m))$ on $\mu$ is shown in Fig.~\ref{fig:temp-ref-sy-end}. As with the data in Fig.~\ref{fig:temp-ref-sy-traj}, we see an insignificant influence of burstiness and edge lifetimes, given the close agreement between the empirical values and the predictions from the PTE and PTN reference models. Indeed, within the small variations in the figure, we find that the extent of consensus predicted by PTE is slightly larger than for the empirical network in the case of the ``Conference'' data set and larger values of $\mu$, and slightly smaller than for the empirical network in the case of the ``High school'' data set for intermediate values of $\mu$. Thus, like epidemic spreading and diffusion dynamics, burstiness can induce both speed-up and slow-down effects~\cite{non-station,diffusion,causa}.

%As convergence parameter $\mu$ is increased, $\Ex(d_m)$ on empirical networks is decreased first and then increased. The minimum of mean opinion distance is achieved around $\mu=0.5$ where the opinion difference between two individuals is eliminated after update (see Eq.~\eqref{eq:dist-up}).

As can be seen in Fig.~\ref{fig:temp-ref-sy-end}, the predicted $\Ex(d(m))$ is reduced significantly for PT compared with PTN. We attribute this enhancement of consensus formation to an extension of nodal lifetimes to the sampling window of the data set. Indeed, the reduction is particularly pronounced in the case of the ``Conference'' and ``Hospital'' data sets for which nodes enter and leave the network dynamically, and more limited in the case of the ``High school'' data set which is well described by the ongoing-node picture. Indeed, the significant reduction that nevertheless occurs for the ``High school'' data set highlights the sensitivity of consensus formation speeds with respect to node lifetimes, as only a small fraction of nodal lifetimes are changed by PT in this data set. This sensitivity can be explained by the concurrency nature of consensus formation. In particular, nodes that do not enter a network at the beginning of the sampling window will cause fluctuation and oscillation of convergence toward consensus, while nodes that leave early have low chances to achieve consensus.

 It is instructive to investigate the expected terminal mean square distance on RT, given by $\Ex(d(m))$ with the updated value of $m$ that results from each application of RT. As seen in  Fig.~\ref{fig:temp-ref-sy-end}, the predictions are very close to those for PT. Both PT and RT extend the lifetimes of nodes to the sampling window. The coincidence between PT and RT again suggests that the rate of consensus formation is mainly determined by the lifetime of nodes rather than the nature of edge and node IETs.

\begin{figure}[!ht]
\centering
\includegraphics[width=1.0\textwidth]{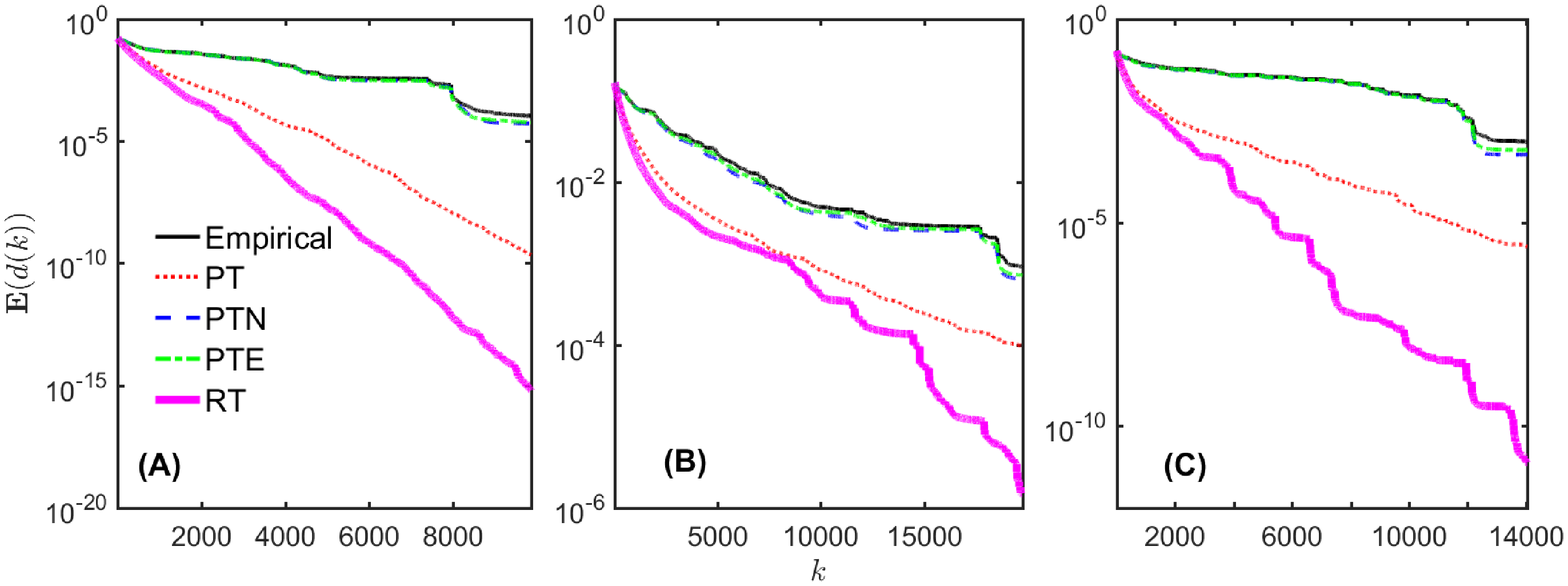}
\caption{Evolution of $\Ex(d(k))$ with $k$ on the empirical networks and their corresponding reference models obtained by applying stochastic Deffuant updates to th graph sequence $\{G_l\}$ with $\mu=0.8$. (A) ``Conference'', (B) ``High school'', and (C) ``Hospital''. The curves for the reference models are averages over 10 realizations.}
\label{fig:temp-ref-asy-traj}
\end{figure}

For the dynamics given by applying stochastic Deffuant updates to the graph sequence $\{G_l\}$, $\Ex(d(k))$ must be obtained by numerical sampling and simulation. Figure~\ref{fig:temp-ref-asy-traj} shows variations in $\Ex(d(k))$ with increasing $k$ in the case that $\mu=0.8$, obtained by averaging over 100 realizations of the initial opinion vector and edge selection sequences. As in the case of the deterministic update model in the previous subsection, the results for the empirical networks match closely with the preditions of the PTE and PTN reference models while PT results in a significant reduction of the expected mean square distance. This is consistent with the observation that the speed of consensus formation is mainly determined by nodal lifetimes. In contrast to the earlier result, RT yields faster decay compared with PT. We attribute this to the likelihood, in the case of PT, that some interactions in a graph will not be selected by the algorithm, while this possibility is essentially eliminated in RT because most graphs have a single interaction. Such missing interactions can slow down consensus formation.

\begin{figure}[!ht]
\centering
\includegraphics[width=1.0\textwidth]{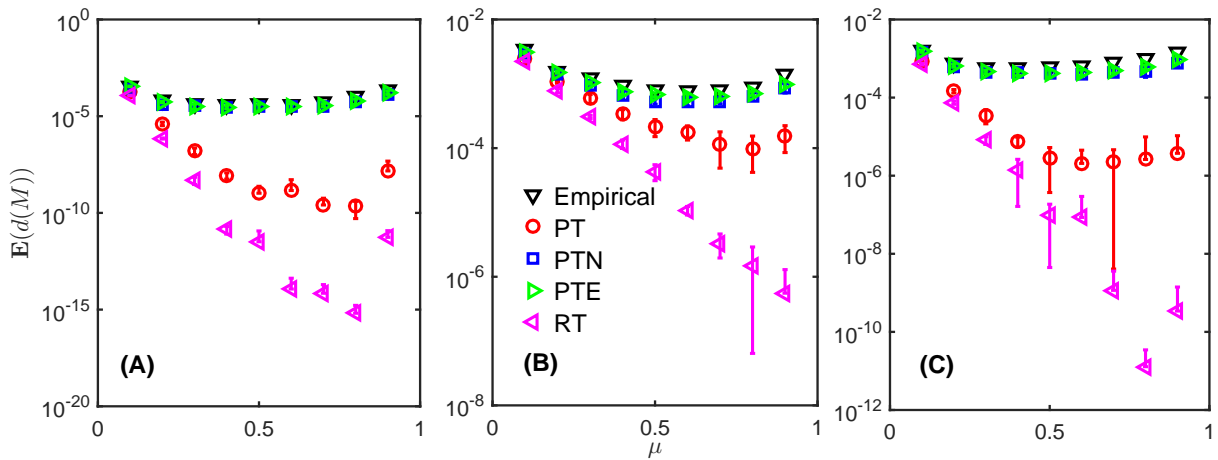}
\caption{The dependence of $\Ex(d(M))$ on $\mu$ for each empirical network and the corresponding reference models using stochastic Deffuant updates applied to the graph sequence $\{G_l\}$. (A) ``Conference'', (B) ``High school'', and (C) ``Hospital''. Results for the PTE, PTN, PT and RT reference models are presented using error bar representing  averages $\pm$ one standard deviation over 10 realizations. In the case that the average minus one standard deviation results in a negative number, we omit the lower half of the error bar.}
\label{fig:temp-ref-asy-end}
\end{figure}

The dependence of $\Ex(d(M))$ on $\mu$ is shown in Fig.~\ref{fig:temp-ref-asy-end}. 
As with the data in Fig.~\ref{fig:temp-ref-sy-end}, the 
differences between empirical networks and reference models are generally amplified as $\mu$ is increased. We again observe insignificant effects of IETs and apparent sensitivity to nodal lifetimes.
 
% add a sentence to talk about the difference between RT and PT

%==================================================%
\section{Further discussions and conclusions}
\label{sec:conclusion}
This paper has explored the effects of  temporal patterns in face-to-face interaction networks on the speed of consensus formation using suitably constructed models of opinion dynamics. Key observations include a slowing down of consensus formation associated with edge weight heterogeneity in models applied to static aggregated topologies, and sensitivity to nodal lifetimes in models applied to graph sequences the preserve temporal ordering. In particular, the extension of nodal lifetimes associated with commonly used randomized reference models can significantly overestimate the speed of consensus formation. Notably, no significant effects were found of burstiness of IETs or edge lifetimes.

Our results provide an important counterpoint to observations in ref.~\cite{accel} on changes to consensus formation associated with PT reference models. The authors conclude that the randomization of contact orders associated with PT can accelerate consensus formation. However, as observed in this paper, changes to contact order that preserve nodal lifetimes, as generated by the PTN reference model, are not associated with accelerated consensus formation. Instead, it appears that the acceleration observed by ref.~\cite{accel} is a consequence of the extension of nodal lifetimes to the full sampling window that results from PT.

We note that Deffuant models are not restricted to opinion dynamics but can be applied to problems of coordination~\cite{accel,multi-agent-05} and synchronization~\cite{syn,dynamical-process}. For example, in \cite{accel}, the original Deffuant model with $\mu=(1-\text{exp}(-2\tau))/2 \in (0,0.5)$ was applied to a problem of continuous-time coordination. We believe that the generalizations developed in this paper, providing for enhanced rates of consensus formation by appropriate tuning of $\mu$  and/or accounting for the effects of group conversations in empirical data sets, can be of similarly significant value to consensus problems in multi-agent coordination, e.g., formation control of mobile robots. Future work should investigate the influence on consensus formation in temporal networks from group conversations.

The confidence bound $\epsilon$ was ignored in our study to simplify the analysis. However, as observed in ref.~\cite{mason}, variations in the time for convergence to steady-state are small for $\epsilon>0.5$, since this typically consists of only one opinion group. We therefore expect that the conclusions of this paper apply also to models that include confidence bounds with $\epsilon>0.5$. For $\epsilon<0.5$, we expect a fragmentation of the opinion dynamics with a steady-state consisting of several distinct opinion groups with consensus within each group~\cite{Deff,mason}. We leave a study of the influence of temporal structure on such fragmentation to future work.

Finally, we note a number of opportunities to generalize the analysis in this paper, for example, by considering heterogeneity of initial opinions and convergence parameters. There is also value in investigating the dependence of the consensus dynamics on outlier nodes~\cite{longlast} with the aim to identify mechanisms for detecting nodes that delay overall convergence.

\section{Acknowledgements}
This material is based upon work supported by the National Science Foundation under Grant No. BCS-1246920.

%%%%%%%%%%%%%%%%%%%%%%%%%%%%%%%%%%%%%%%%%%%%%%%%%%%%
\appendix

%================================================%
\section{Row-stochastic matrices}
\label{sec:deffuant}

Let $\{w^{(k)}\}_{k=1}^n\in\mathbb{R}^{N\times N}$ be a sequence of diagonal matrices and $\{L^{(k)}\}_{k=1}^n\in\mathbb{R}^{N\times N}$ be a sequence of Laplacian matrices, such that
\begin{equation}
    \sum_{j=1}^N L_{ij}^{(k)}=0,\,\forall (i,k)\in\{1,\ldots,N\}\times\{1,\ldots,n\}
\end{equation}
and such that the elements of $I-w^{(k)}L^{(k)}$ lie in the interval $[0,1]$. The following lemma then follows by induction.
%------------------------------------------%
\begin{lemma}
\label{lemma_1}
The matrix 
\begin{equation}
    T = \prod_{k=1}^n \left(I-w^{(k)}L^{(k)}\right)
\end{equation}
%\begin{equation}
%\label{eq:T2}
%    T^{(ab)} = \left(I-w^{(a)}L^{(a)}\right)\left(I-w^{(b)}L^{(b)}\right)
%\end{equation}
is a row-stochastic matrix, i.e., $\sum_{j=1}^N T_{ij}=1, \forall i\in\{1,...,N\}$ and the elements of $T$ lie in the interval $[0,1]$.
\end{lemma}

Using the result of this lemma, it is straightforward to conclude that the expectations for the transition matrices associated with each of the four models of opinion dynamics on time-dependent networks derived in section~\ref{sec:timedep networks} are also row-stochastic matrices.

%\section*{\refname}
%\bibliographystyle{elsarticle-num} 
%\bibliography{bbl}


%\begin{thebibliography}{00}



%\end{thebibliography}

\end{document}